\def\aa{A\&A} %Astronomy & Astrophysics%
\def\aasup{A\&AS} %A & A Supplements%
\def\anj{AJ} %Astronomical Journal%
\def\apj{ApJ} %Astrophysical Journal%
\def\mn{MNRAS} %Monthly Notices of the Royal...%
\def\rev{ARA\&A} %Annual Review of Astronomy & Astrophysics%
\def\nar{NewAR} %New Astronomy Reviews%
\def\phe{\phantom{1}}
\def\ph2{\phantom{88}}
\def\ph.8{\phantom{.8}}%
\documentclass{aa}
\begin{document}

\thesaurus{11.09.1 8C 0821+695; 11.01.2; 11.03.3; 13.18.1} 

\title{The giant radio galaxy 8C\,0821+695 and its environment}

\author{L. Lara\inst{1} \and 
K.-H. Mack\inst{2,3}  \and M. Lacy\inst{4} \and  U. Klein\inst{3} \and
W.D.      Cotton\inst{5,6}    \and     L.      Feretti\inst{2}    \and
G. Giovannini\inst{2,7} \and M. Murgia\inst{2,8}}
\offprints{L. Lara}

\institute{Instituto de Astrof\'{\i}sica de Andaluc\'{\i}a (CSIC),
Apdo. 3004, 18080 Granada, Spain
\and 
Istituto  di  Radioastronomia  (CNR),  Via P.   Gobetti  101,  I-40129
Bologna, Italy
\and
Radioastronomisches Institut  der Universit\"at Bonn,  Auf dem H\"ugel
71, D-53121 Bonn, Germany
\and
IGPP,  L-413, Lawrence  Livermore National  Laboratory,  Livermore, CA
94550, USA
%Astrophysics, Department of Physics, Keble Road, Oxford OX1 3RH (England)
\and
National   Radio    Astronomy   Observatory,   520    Edgemont   Road,
Charlottesville, VA 22903-2475, USA
\and 
Sterrewacht Leiden, Niels Bohrweg 2, 2300-RA Leiden, The Netherlands
\and
Dipartimento di  Fisica, Universit\`a di Bologna, Via  B.  Pichat 6/2,
I-40127 Bologna, Italy
\and
Dipartimento di  Astronomia, Universit\`a  di Bologna, Via  Ranzani 2,
I-40127 Bologna, Italy }

\date{Received / Accepted}

\authorrunning{Lara et al.} 
\titlerunning{GRG 8C\,0821+695 and its environment}

\maketitle

\begin{abstract}

We present  new VLA  and Effelsberg observations  of the  radio galaxy
8C\,0821+695. We have obtained  detailed images in total intensity and
polarization  of this 2  Mpc sized  giant.  The  magnetic field  has a
configuration  predominantly parallel  to  the source  main axis.   We
observe Faraday rotation at low frequencies, most probably produced by
an ionized  medium external  to the radio  source. The  spectral index
distribution is  that typical of  FR II radio galaxies,  with spectral
indices  gradually steepening  from  the source  extremes towards  the
core. Modeling  the spectrum in  the lobes using  standard synchrotron
loss  models  yields the  spectral  age of  the  source  and the  mean
velocity  of the  jet-head with  respect  to the  lobe material.   The
existence of a  possible backflow in the lobe  is considered to relate
spectral  with dynamical determinations  of the  age and  the velocity
with respect to  the external medium. Through a  very simple model, we
obtain a physical characterization of the jets and the external medium
in which  the radio galaxy  expands.  The results in  8C\,0821+695 are
consistent with  a relativistic jet nourishing the  lobes which expand
in a  hot, low  density halo.  We infer a  deceleration of  the source
expansion velocity which we  explain through a progressive increase in
the hot-spot size.

\keywords{galaxies: individuals (8C\,0821+695)
          galaxies: active -- intergalactic medium -- radio continuum:
          galaxies }
\end{abstract}

\section{Introduction}

Giant  Radio Galaxies  (GRGs)  constitute an  unusual  class of  radio
sources  with projected  linear  sizes larger  than 1  Mpc\footnote{We
assume  H$_{0}$=75  $  \mathrm{km\,s^{-1}\,Mpc^{-1}}$ and  q$_{0}=0.5$
throughout this paper.}. These objects, although giant in size, do not
stand out in luminosity, most  of them being of low surface-brightness
(Subrahmanyan et al.  \cite{subra2}).   Moreover, the enormous size of
GRGs sometimes hampers the identification of emission from two distant
lobes as  part of  an individual object  (e.g.  B2~1358+305;  Parma et
al. \cite{parma}).  Probably, these are  the reasons why to date there
are only about 50 known GRGs (Ishwara-Chandra \& Saikia
\cite{ishwara}; Schoenmakers et al. \cite{arno1}).  Only recently, 
radio surveys like the Northern VLA Sky Survey (NVSS; Condon et al.
\cite{nvss}) or the Westerbork Northern Sky Survey (WENSS; Rengelink
et al.   \cite{wenss}) allow sensitive searches of  GRGs with adequate
angular resolution.

Two  basic scenarios have  been envisaged  to explain  the outstanding
sizes of GRGs.  First, their  lobes could be fed by extremely powerful
central engines which  would endow the jets with  the necessary thrust
to bore their long way  through the ambient medium. Second, GRGs could
be  normal radio  sources  evolving in  very low-density  environments
offering little  resistance to the  expansion of the jets.   While the
first  possibility  requires  the  existence of  prominent  cores  and
hot-spots, which are not always observed (Ishwara-Chandra \& Saikia
\cite{ishwara}), the second possibility is supported by the high degree 
of polarization found in GRGs also at low frequencies (Willis \& O'Dea
\cite{willis}) and  seems to  be the most  plausible scenario  in most
cases (Mack et al.  \cite{mack97b}).  It seems that the expansion of a
radio galaxy in a low density environment during a time long enough to
allow reaching Mpc  sizes, rather than higher than  usual radio powers
or expansion velocities, are the two basic ingredients to build up the
GRG population (Schoenmakers et al. \cite{arno2}).

GRGs are located in regions hardly accessible via direct observations:
they  do   not  reside  in  rich  galaxy   clusters  (Subrahmanyan  et
al. \cite{subra2})  and the X-ray emission around  their host galaxies
is usually  weak (Mack et al.   \cite{mack97a}).  However, information
about  the  ambient medium  at  very  large  distances from  the  host
galaxies  can  still  be  gained  through the  study  of  their  radio
properties.  Most GRGs probe  the intergalactic medium (IGM) providing
information about the density of matter outside galactic halos, adding
important  observational constraints  to  current cosmological  models
(Begelman \& Cioffi \cite{begelman}; Nath \cite{nath}).

We discuss in  this paper new observations of  the GRG 8C\,0821+695, a
Fanaroff-Riley  type II  (Fanaroff \&  Riley \cite{fanaroff}).   It is
optically  identified  with  a  faint  $M_{R}\sim 22.2$  galaxy  at  a
redshift  of 0.538  (Lacy et  al.  \cite{lacy}).   Its  high redshift,
compared to other GRGs, renders 8C\,0821+695 a very interesting object
since  it provides information  on the  external environment  at large
cosmological  distances.  No  X-ray source  coincident with  the radio
source is  found in the Bright  Point Source catalogue  from the ROSAT
All  Sky  Survey.  At  the  distance  of  8C\,0821+695, one  arcsecond
corresponds to 4.9 kpc.

\section{Observations and data analysis}

We observed 8C\,0821+695  with the VLA in the  framework of a complete
sample of large angular size radio sources selected from the NVSS (see
Lara et al. \cite{lara} for  a sample description), and with the 100-m
Effelsberg telescope. We also incorporate for the discussion maps from
the   NVSS  and   WENSS,   and   maps  presented   by   Lacy  et   al.
(\cite{lacy}). The data have been calibrated according to the scale of
Baars et al. (\cite{baars}).

\subsection{WENSS and NVSS maps}

8C\,0821+695 appears  as a straight  $\sim$7\arcmin\ long FR  II radio
source  in the  NVSS  and the  WENSS maps,  with  its main  axis at  a
position angle (P.A.) of 11\degr, measured north through east.

\begin{figure*}
\vspace{9cm}
\includegraphics{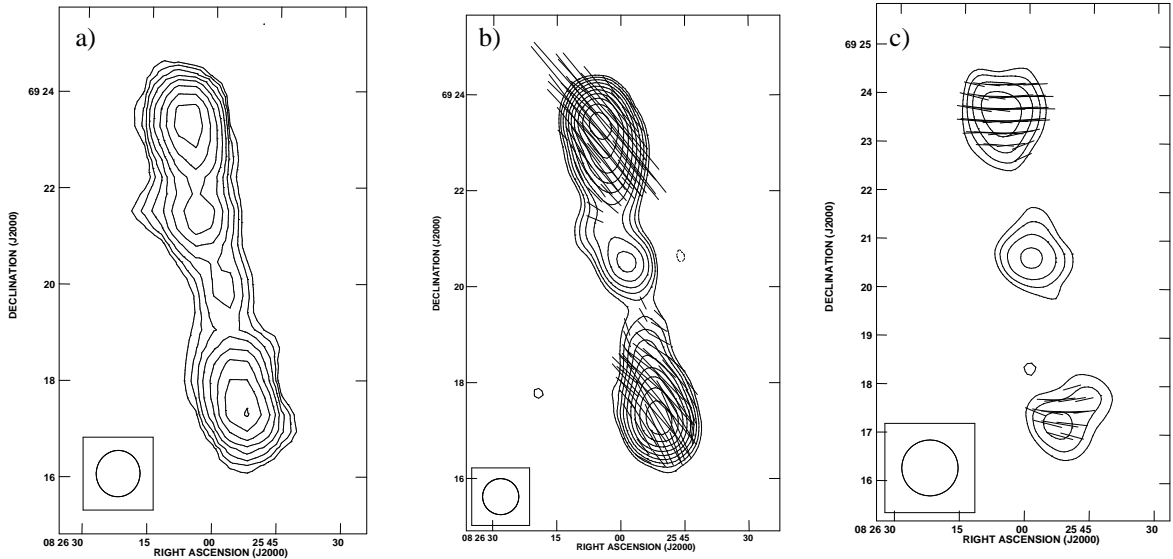}
%\rule{0.4pt}{4cm}% line thickness, height of picture
\caption{
Low resolution  maps of  8C\,0821+695 from {\bf  a)} WENSS  (327 MHz),
{\bf b)}  NVSS (1400 MHz)  and {\bf c)} Effelsberg  observations (10.6
GHz). Contours are spaced by factors of $\sqrt{2}$ in brightness, with
the lowest at  3 times the rms noise  level.  The superimposed vectors
represent  the polarization  position angle  (E-vector),  with lengths
proportional to  the amount  of polarization. From  left to  right, we
list below the  rms noise level, the equivalence  of $1\arcsec$ length
in polarized intensity and the  Gaussian beam used in convolution: rms
= 3.3,  0.45 and 0.5 mJy  beam$^{-1}$; $1\arcsec$ = 33  and 20 $\mu$Jy
beam$^{-1}$;  Beam  =  $57\farcs7  \times  54\farcs0$  P.A.  $0\degr$,
$45\arcsec \times 45\arcsec$ and $69\arcsec\times69\arcsec$.  }
\label{lowres}
\end{figure*}

The WENSS map  (Fig.~\ref{lowres}a), at a frequency of  327 MHz and an
angular   resolution  of   57\farcs7$\times$54\farcs0,   presents  two
prominent  lobes  (N  and  S)  connected by  a  continuous  bridge  of
emission, although the position of the core is not evident at all.

The NVSS map (Fig.~\ref{lowres}b), made at a frequency of 1400 MHz and
an  angular resolution of  45\arcsec, shows  a prominent  central core
straddling  the  two  radio  lobes.   The N-lobe  has  higher  surface
brightness  than   the  S-lobe.   The   mean  fractional  polarization
($p_{m}$) at 1400  MHz is $p_{m}$=23\% in the  N-lobe and $p_{m}$=26\%
in the  S-lobe, while the core  is unpolarized.  The  E-vectors have a
similar and rather uniform orientation  in the N- and S-lobes, oblique
to the source main axis.

\subsection{Effelsberg observations}

We observed  8C\,0821+695 with the 100-m Effelsberg  telescope at 10.6
GHz  (Tab.~\ref{obs})  in  order   to  obtain  information  about  the
morphology  and  polarization  properties  at high  frequencies.   The
observational  and data  reduction procedures  were those  detailed by
Gregorini  et  al.   (\cite{gregorini}).   The maps  were  CLEANed  as
described by  Klein \& Mack (\cite{kleinmack}).  We  made 40 coverages
resulting --after combining--  in a final noise level  of 0.5 mJy/beam
in  total power  and  0.1  mJy/beam in  the  polarized channels.   The
polarization   maps  were   corrected  for   the   non-Gaussian  noise
distribution  of the polarized  intensity, as  described by  Wardle \&
Kronberg (\cite{wardle}).  This is of particular importance in case of
polarized low-brightness regions, e.g. in extended radio lobes.
 
The 10.6  GHz map (Fig.~\ref{lowres}c), with an  angular resolution of
$69\arcsec$, shows three components  corresponding to the core and the
two  lobes.    The  superimposed  vectors   represent  the  electrical
fields. Since at this high frequency Faraday effects are most probably
negligible, a rotation by 90\degr~ immediately yields the direction of
the projected  magnetic field.  It is  oriented predominantly parallel
to the source main axis.  The degree of polarization at this frequency
is 26\% in both lobes.

We also  observed 8C\,0821+695 with  the 100-m telescope at  4850 MHz.
After combination  of 10 coverages  we reached the confusion  limit of
0.6  mJy/beam  in total  intensity,  while  the  noise level  was  0.1
mJy/beam in the polarization channels.  Because of the large beam size
(143\arcsec), this  map does  not reveal any  additional morphological
details,  so  we do  not  need  to display  it  here.   The degree  of
polarization derived  from the Effelsberg map  at 4850 MHz  is 19\% in
the N- and S-lobes.

\begin{table}[b]
\caption[]{Observations of 8C\,0821+695}
\label{obs}
\begin{tabular}{lrrcr}
\hline
Instrument & $\nu$ & $\Delta\nu$ &  Duration & Date \\ & (MHz) & (MHz)
           & (min) & \\
\hline
NVSS & 1400  & 100 & -- &  23 Nov 93 \\ WENSS  & 327 & 5 & --  & -- \\
VLA-C & 1425 & 50 & 10 & 19 Feb 96 \\ & 4860 & 100 & 10 & 19 Feb 96 \\
VLA-B & 1425 & 50 & 10 & 19 Nov 95 \\ & 4860 & 100 & 10 & 25 May 97 \\
Effelsberg & 10550  & 300 & -- & 28 Aug  94 \\ & 4850 & 500  & -- & 27
Feb 98 \\
\hline 
\end{tabular}
\end{table}

\subsection{VLA observations}

We made continuum observations of  8C\,0821+695 with the VLA in the B-
and C-configurations  at 1425 and 4860 MHz in dual polarization (see Tab.~\ref{obs}).  The
interferometric  phases   were  calibrated  with   the  nearby  source
J0903+679, except  during 4860 MHz observations with  the B-array, for
which J0841+708 was used as  phase calibrator. The radio sources 3C286
and 3C48 served as primary  flux density calibrators.  Data from the B
and C  arrays were combined in  order to take advantage  of the higher
B-array resolution  and of the higher C-array  sensitivity to extended
emission.  The  calibration and mapping  of the data were  carried out
with the NRAO AIPS package. Maps  at 4860 MHz had to be corrected from
primary beam attenuation. In addition, correction for the non-Gaussian
noise distribution in the polarized intensity map was applied.

\begin{figure*}
\vspace{13cm}
%\rule{0.4pt}{4cm}% line thickness, height of picture
\includegraphics{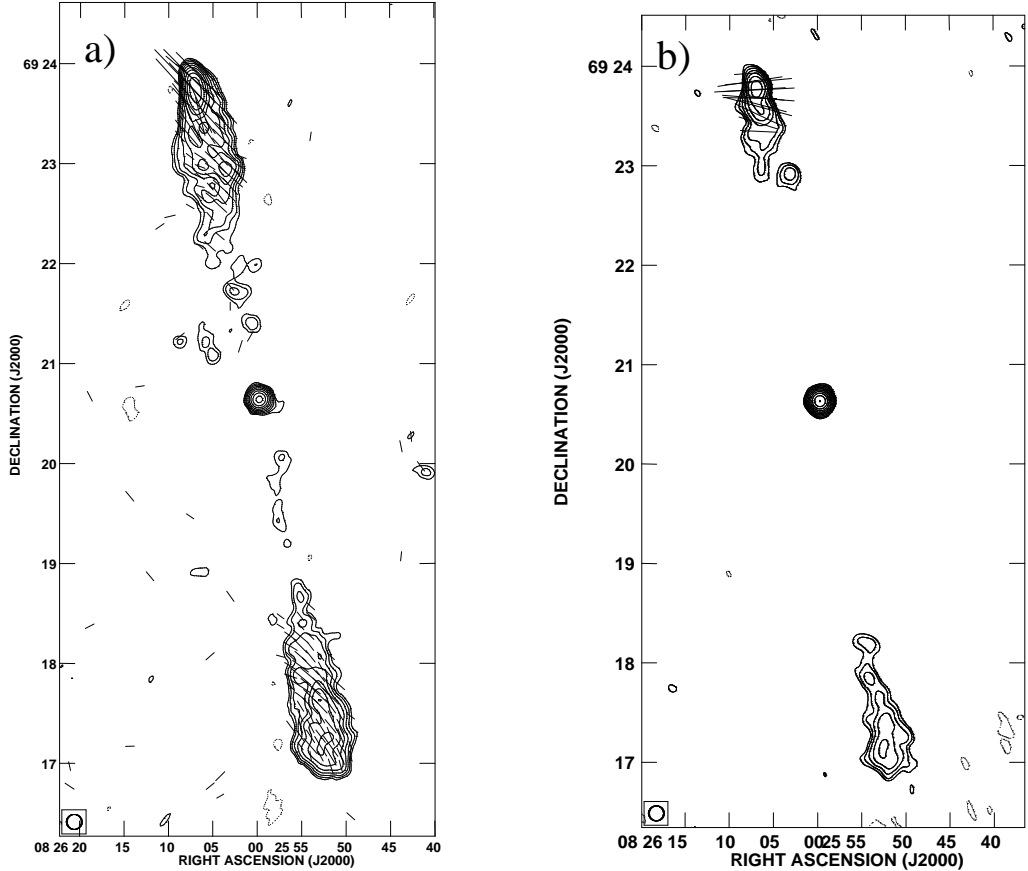} 
\caption{
{\bf a-b} VLA B+C-array maps of  8C\,0821+695 at 1425  (left) and 4860
MHz  (right).   Vectors  represent  the  polarization  position  angle
(E-vector),  with length  proportional to  the amount  of polarization
(1\arcsec~ corresponds to  35 $\mu$Jy beam$^{-1}$ on the  1425 MHz map
and to  12.5 $\mu$Jy beam$^{-1}$ on  the 4860 MHz  map).  Contours are
spaced by  factors of $\sqrt{2}$ in  brightness, with the  lowest at 3
times  the   rms  level   (rms  =  0.13   and  0.1   mJy  beam$^{-1}$,
respectively).  The Gaussian  beam size is 9\arcsec$\times$9\arcsec~in
both maps.
} 
\label{vla}
\end{figure*}

The VLA  map at 1425  MHz shows a  well defined compact core,  and two
lobes of extended emission  (Fig.~\ref{vla}a). The N-lobe is dominated
by a compact  component, suggesting the existence of  a faint hot-spot
at the end of the jet, while  a similar feature is not observed in the
S-lobe.   The polarized  emission of  8C\,0821+695 at  1425  MHz comes
predominantly from this hot-spot in the N-lobe. At  this position, the
degree of polarization is 30\%.  It  is 20\% in the rest of the N-lobe
and 25\% in the S-lobe.  As  in the NVSS map, the electric vectors are
inclined with respect to the source main axis, probably due to Faraday
rotation.

The  VLA  map   of  8C\,0821+695   at   4860  MHz   is  displayed   in
Fig.~\ref{vla}b.  At  this frequency, the  core is the  most prominent
feature.   The  hot-spot  in the  N-lobe  appears  at  the end  of  an
elongated  structure aligned with  the radio  axis.  The  N-lobe shows
clear  oscillations in  its ridge  line  that are  reminiscent of  the
``dentist drill''  model which assumes a  jitter of the  jet-head with
time (Scheuer \cite{scheuer}).  On  the other hand, the S-lobe appears
much more  ``relaxed'' and  without a strong  hot-spot.  We  measure a
total   source  length  of   6\farcm95  from   Fig.~\ref{vla}b,  which
corresponds to a  projected linear size of 2  Mpc.  The north-to-south
arm ratio is 0.89.

The polarized emission  of 8C\,0821+695 at 4860 MHz  is again dominated
by the N-lobe hot-spot, where we measure a mean degree of polarization
of  22\%, reaching  26\%  in  the brightest peak. E-vectors  are
predominantly oriented perpendicularly to the source main axis.  We do
not detect  significant polarization in  the S lobe at  this frequency
and resolution.

Our highest resolution observations  (VLA B-array at 4860 GHz) provide
the  following  coordinates  for   the  compact  core  of  8C\,0821+695
(J2000.0):  RA  =  08$^h$25$^m$  59\fs770, DEC  =  +69\degr  20\arcmin
38\farcs59, fully consistent with the position of the host galaxy.

\section{Results}

\subsection{Rotation Measure and depolarization}

We have estimated the Rotation Measure ($RM$) over the two radio lobes
of 8C\,0821+695.  To do that, we  convolved the NVSS 1400  MHz and the
VLA 4860 MHz polarization maps to  the beam of the Effelsberg 10.6 GHz
map (a  circular Gaussian beam  of $69\arcsec$ FWHM), and  applied the
AIPS task  RM.  Although the  convolution of the  VLA map with  such a
large beam is  in general not advisable, in  this case the orientation
of the polarization vectors was not seriously affected. We did not use
the Effelsberg 4850 MHz map because of its too low angular resolution,
which would prevent us from finding any possible structure in the $RM$
distribution.  Even so,  we obtain a rather uniform  $RM$ over the two
lobes of 8C\,0821+695, with an average value of $-20$ rad~m$^{-2}$.

The uniform and similar distributions of the electric field P.A. and of
the $RM$ over the two lobes suggest that the contribution of the radio
source  to  the  observed  $RM$  is negligible.   Moreover,  the  mean
galactic $RM$ at the position of 8C\,0821+695 (galactic coordinates $l=
145\fdg7$ and $b= 33\fdg54$) lies between $-30$ rad~m$^{-2}$ and 0
rad~m$^{-2}$ (Simard-Normandin  \& Kronberg \cite{simard}), consistent
with our observed  value.  We thus conclude that  the observed Faraday
rotation is most plausibly of  galactic origin, although a small local
halo contribution between $-20$ to 10 rad~m$^{-2}$ cannot be excluded.

The  fractional  polarizations  derived  at different  frequencies  are
essentially  consistent  with  the   lack  of  depolarization  at  low
frequencies. This  is clearly derived  from the comparison of  the low
resolution data (10.6 and 5 GHz  Effelsberg data and 1.4 GHz data from
NVSS), taking into account that the 5 GHz data have a much larger beam
and therefore are likely to  suffer from beam depolarization. The higher
resolution VLA data show that the two lobes are still highly polarized
at 1.4 GHz, in agreement with the presence of a magnetic field ordered
on  the  restoring  beam  scale.   We ascribe  the  lack  of  detected
polarization in the S-lobe at 5 GHz,  with 9\arcsec~resolution, to the
lower sensitivity of this image to extended structure.

\subsection{The broad-band radio-spectrum of 8C\,0821+695}

\begin{figure}
\vspace{7cm}
\includegraphics{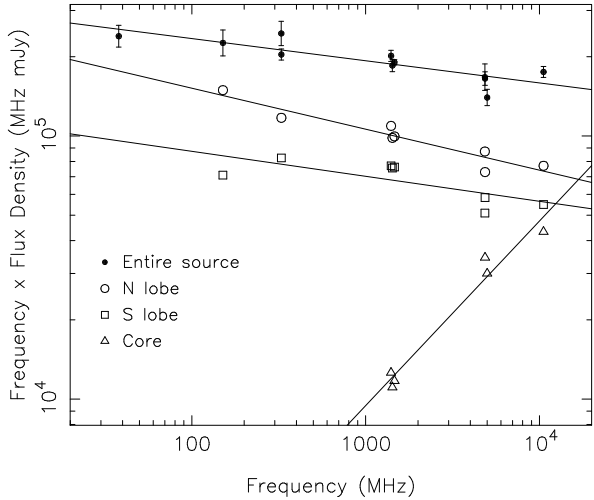} 
%\rule{0.4pt}{4cm}
\caption{Radio spectrum of 8C\,0821+695 and its
components. See
Table~\ref{flux} for numerical values and their references.} 
\label{espectro}
\end{figure}

We  have   compiled  all   available  flux  density   measurements  on
8C\,0821+695 and  have plotted them  as a function of  frequency giving
values, when  possible, for  the whole source,  the two lobes  and the
core  separately   (Tab.~\ref{flux};  Fig.~\ref{espectro}).   We  have
obtained  the  spectral index $\alpha$  (defined  so  that the  flux
density $S \propto \nu^{-\alpha}$) of the different components from linear 
fits to the data:  $\alpha_N=1.15\pm0.02$; $\alpha_S=1.10\pm0.04$.   
On  the other hand,  the core  shows  a flat  spectrum  with
a mean $\alpha = 0.30\pm 0.07$.
%below 4.9  GHz  
%and a  steep one at higher frequencies.

\begin{table}[b]
\caption[]{Flux density of 8C\,0821+695}
\label{flux}
\begin{tabular}{rr@{$\pm$}rrrcc}
\hline
\multicolumn{1}{c}{Freq.} & \multicolumn{5}{c}{Integrated flux density of}    & Ref.  \\
          & \multicolumn{2}{c}{Entire source} & N lobe & S lobe & Core &            \\ 
\multicolumn{1}{c}{(MHz)} & \multicolumn{2}{c}{(mJy)}      &\multicolumn{1}{c}{(mJy)}&\multicolumn{1}{c}{(mJy)}& \multicolumn{1}{c}{(mJy)}&            \\
\hline
38   &6300\ph.8 &600\ph.8 &\multicolumn{1}{c}{--}&\multicolumn{1}{c}{--}& --     & (1) \\
151  &1495\ph.8 &170\ph.8 &989\ph.8              &470\ph.8              & --     & (2) \\
327  &750\ph.8  &80\ph.8  &\multicolumn{1}{c}{--}&\multicolumn{1}{c}{--}& --     & (3) \\
327  &623\ph.8  &31\ph.8  &358\ph.8              &252\ph.8              & --     & (4) \\
1400 &144\ph.8  &7\ph.8   &78\ph.8               &55\ph.8               & 9\ph.8 & (5) \\
1425 &130\ph.8  &7\ph.8   &69\ph.8               &53\ph.8               & 8\ph.8 & (6) \\
1465 &130\ph.8  &3\ph.8   &68\ph.8               &52\ph.8               & 8\ph.8 & (2) \\
4850 & 34.5     &4\ph.8   &18\ph.8               &10.5                  & --     & (7) \\
4860 & 34\ph.8  &2\ph.8   &15\ph.8               &12\ph.8               & 7\ph.8 & (6) \\
5000 & 28\ph.8  &2\ph.8   &\multicolumn{1}{c}{--}&\multicolumn{1}{c}{--}& 6\ph.8 & (2) \\
10550& 16.6     &0.8      &7.3                   &5.2       & 4.1    & (7) \\
\hline 
\end{tabular}
\begin{list}{}{}
\item[](1) Rees 1990; (2) Lacy et al. 1993; (3) Wieringa 1991; 
\item[](4) WENSS; (5) NVSS; (6) This work, VLA; 
\item[](7) This work, Effelsberg
\end{list}
\end{table}

\begin{figure}
\vspace{7cm}
\includegraphics{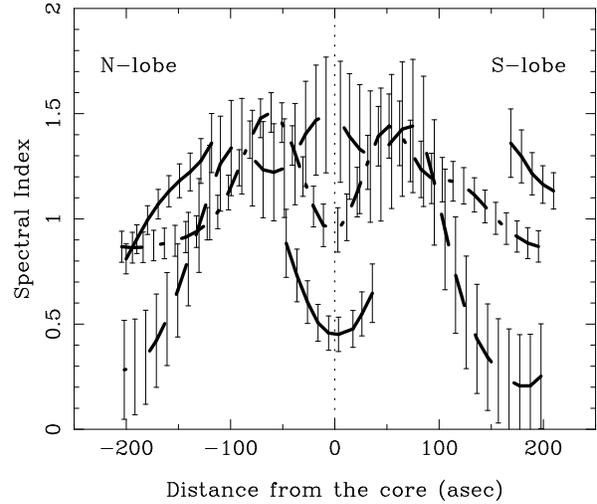} 
%\rule{0.4pt}{4cm}
\caption{Profiles of the spectral index of 8C\,0821+695 along its main axis 
with an angular resolution of 69\arcsec. 
$- - -$: spectral index between 151 and 327 MHz; 
$-\cdot-\cdot-$: spectral index between 327 and 1400 MHz; 
\line(1,0){25} : spectral index between 1400 and 10550 MHz.} 
\label{index}
\end{figure}

In  order  to study  the  dependence  of  the spectral  properties  of
8C\,0821+695 with frequency  and distance from the core,  we have made
three  low   resolution  spectral  index  maps  using   the  10.6  GHz
Effelsberg, the 1400 MHz NVSS, the 327 MHz WENSS and the 151 MHz CLFST
map by Lacy et al.~(\cite{lacy}).  To construct the $\alpha$-maps, all
total intensity maps were convolved to a circular beam of $69\arcsec$,
and  then  were  registered  to  the nominal  position  of  the  core.
Figure~\ref{index}  displays spectral  index profiles  along  the main
radio axis for the  three frequency intervals, with plotted 1-$\sigma$
errors deduced taking  into account the rms noise  level of each image
and the uncertainties in their flux density scales.  The core shows up
as a flattening of the spectrum  at the center of the higher frequency
profiles.  In both lobes we find an overall steepening of the spectrum
from the extremes towards the  compact core, a behaviour typical of FR
II radio  sources. In  addition, the steepening  of the  spectrum with
increasing frequency  is also evident from this  plot. The spectrum
at low frequencies is rather  flat in the source extremes. This result
is expected  since these  regions are dominated  by the  flat spectrum
hot-spot like  regions and is  consistent with the  injection spectral
index derived in Sect.~3.4.

\begin{figure}
\vspace{12.5cm}
\includegraphics{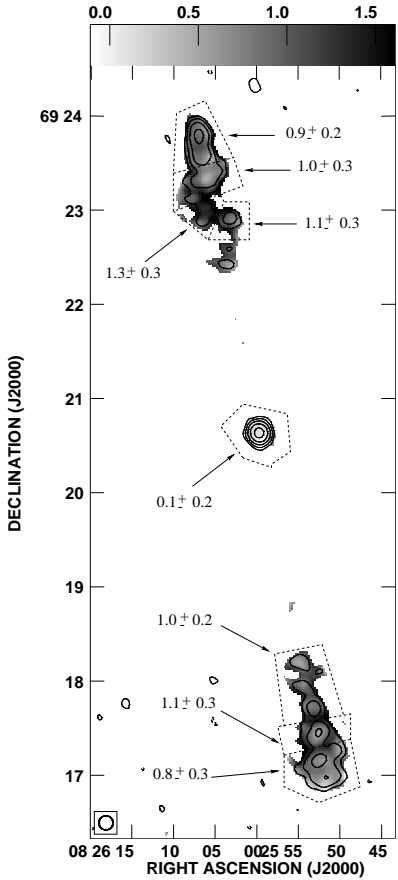}
%\rule{0.4pt}{4cm}
\caption{Spectral index, in grey-scale,  of 8C\,0821+695 between 
4860 and 1425 MHz, with an angular resolution of 9\arcsec$\times$9\arcsec. 
The contours
represent total intensity at 5 GHz to facilitate the identification of
emission regions in the radio source with spectral trends. Numerical values
of the mean spectral index over selected regions are indicated for clarity.} 
\label{alfamap}
\end{figure}

To study in detail the spectral index distribution over 8C\,0821+695 we
have made a high-resolution $\alpha$-map registering our VLA  maps at 1425
and 4860  MHz to the  position of the  core and performing a  
pixel-by-pixel evaluation  of the spectral index between  these two frequencies
(Fig.~\ref{alfamap}).  We find  the  same general  trends observed  at
lower resolutions, but  convolved with a now evident  structure in the
$\alpha$-distribution.  The  core shows a flat  spectrum between these
two frequencies. The lobes  present similar spectral indices, although
the N-lobe  hints at a slightly  steeper spectrum which,  if real, would
indicate a more efficient energy dissipation in this region.

\subsection{Physical parameters of 8C\,0821+695}

\begin{table*}[]
\caption[]{Physical parameters for the N- and S- lobes}
\label{param}
\begin{tabular}{cccccc}
\hline
Distance & FWHM  & Brightness & B$_{me}$  & u$_{me}$  & P$_{eq}$     \\
  (kpc)  & (kpc) &(mJy/beam$^a$)  &($\eta^{-2/7}\mu$G)& ($\eta^{-4/7}$J m$^{-3}$) 
& ($\eta^{-4/7}$N $m^{-2}$)\\ 
\hline
\multicolumn{6}{c} {North lobe} \\
\hline
 {\phe}340 & {\phe}85 & {\phe}2.5 &{\phe}3.8 & 1.4$\times 10^{-13}$ & 8.5$\times 10^{-14}$ \\ 
 \phe 480 & \phe 98 & \phe 1.9 &\phe 3.4 & 1.1$\times 10^{-13}$ & 6.7$\times 10^{-14}$ \\
 \phe 630 & 109    & \phe 4.6 &\phe 4.3 & 1.7$\times 10^{-13}$ & 1.0$\times 10^{-13}$ \\
 \phe 775 & 104    & \phe 7.3 &\phe 4.9 & 2.3$\times 10^{-13}$ & 1.4$\times 10^{-13}$ \\
 \phe 923 & \phe 27 & 42.6    &12.0    & 1.3$\times 10^{-12}$ & 8.2$\times 10^{-13}$ \\
\hline 
\multicolumn{6}{c} {South lobe} \\
\hline
 \phe 433 & \phe 91 &  {\phe}0.8 & {\phe}2.7 & 6.9$\times 10^{-14}$ & 4.3$\times 10^{-14}$ \\
 \phe 579 & \phe 85 &  {\phe}1.4 & {\phe}3.3 & 9.9$\times 10^{-14}$ & 6.1$\times 10^{-14}$ \\
 \phe 726 & \phe 98 &  {\phe}3.1 & {\phe}3.9 & 1.4$\times 10^{-13}$ & 8.9$\times 10^{-14}$ \\
 \phe 872 & \phe 79 &  {\phe}6.8 & {\phe}5.2 & 2.5$\times 10^{-13}$ & 1.6$\times 10^{-13}$ \\
    1019  & 104     &  {\phe}8.8 & {\phe}5.2 & 2.5$\times 10^{-13}$ & 1.6$\times 10^{-13}$ \\
\hline \\
\end{tabular}
\begin{list}{}{}
\item[$^a$] The beam is a circular Gaussian of 15\arcsec$\times$15\arcsec
\end{list}
\end{table*}

We  have measured  the  FWHM  and the  surface  brightness at  several
positions on  the ridge  line of 8C\,0821+695  by fitting  Gaussians to
surface  brightness profiles  taken perpendicular  to the  source main
axis  in   the  high-sensitivity  1400   MHz  map  by   Lacy  et  al.
(\cite{lacy}).  The deconvolution of  the width and surface brightness
was done following Appendix A in Killeen et al.  (\cite{killeen}).  We
used  the  standard formulae  of  synchrotron  radiation (e.g.   Miley
\cite{miley})  to calculate  the  minimum energy  density $u_{me}$  at
these positions  and the corresponding magnetic  field $B_{me}$, which
is  approximately  the equipartition  field.   The  total pressure  is
assumed  to be that  of equipartition  between particles  and magnetic
field, $P_{eq}=0.62~u_{me}$.  Besides,  the following assumptions were
made in the calculations: {\em i)} the magnetic field is assumed to be
random; {\em  ii)} the  energy of particles  is equally stored  in the
form of  relativistic electrons and heavy particles;  {\em iii)} lower
and  upper  frequency  cutoffs  were  set  to  10  MHz  and  100  GHz,
respectively; {\em  iv)} the spectral index  is 1.1; and  {\em v)} the
line-of-sight depth is equal to the deconvolved FWHM.  The results are
listed in Tab.~\ref{param} as a  function of the filling factor of the
emitting regions $\eta$.

\subsection{Spectral aging}

Based  on  the maps  at  151~MHz  (Lacy  et al.~\cite{lacy}),  327~MHz
(WENSS), 1.4~GHz  (NVSS), 4.8~GHz (VLA) and  10.4~GHz (Effelsberg), we
have  performed a spectral  aging analysis  of 8C\,0821+695.   We have
determined the spectrum at different positions of the source averaging
the flux  density over  selected regions with  a beam  equivalent area
($69\arcsec$)  in  order  to  insure independent  measurements.  Three
selected regions were  centered on the N-lobe (at  the position of the
hot-spot and at 711 and 364 kpc  from the core) and other three on the
S-lobe (at the southern extreme and at 903 and 553 kpc from the core).

\begin{figure}
\vspace{10cm}
\includegraphics{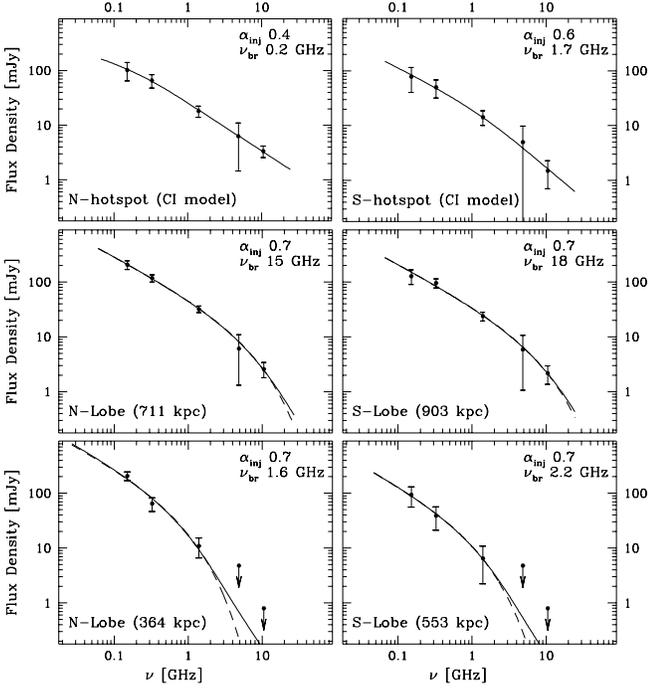}
%\rule{0.4pt}{4cm}

\caption{Spectra at different regions of 8C\,0821+695. Left and right panels 
refer to the N- and S-lobe, respectively. Upper panels correspond to the 
spectra at the  source extremes  (hot-spots), fitted by  CI model.  Middle and
bottom panels correspond to lobe  regions at 711 and 364 kpc (N-lobe),
and 903 and  553 kpc (S-lobe) from the  core, respectively, where data have
been fitted with  a KP (continuous line) and  JP (dashed line) models.
Upper limits in the bottom panels are at the rms noise level of the 
corresponding maps.}

\label{spectra}
\end{figure}

We show  in Fig.~\ref{spectra} the  spectra of the  different selected
regions and the fits to  the data after the application of synchrotron
loss  models.  Left  and  right panels  refer  to the  N- and  S-lobe,
respectively.   The spectra at  the source  extremes (top  panels) are
best fitted by the continuous injection model (CI; Pacholczyk
\cite{pacholczyk}), giving an injection spectral index of $\alpha_{inj}
= 0.4$  for the northern  hot-spot and $\alpha_{inj}  = 0.6 $  for the
southern extreme  of the source,  in agreement with the  low frequency
spectral index in Fig.~\ref{index}.  The spectra in these regions show
a break at  low frequencies ($\sim 1$ GHz)  with a moderate steepening
afterwards.   However,  the flat  spectrum  of  the  hot-spots at  low
frequencies might indicate that the source here is optically thick and
any   spectral   fit   in   these   regions   must   be   taken   with
caution. Moreover, Meisenheimer et al. (\cite{meisenheimer}) find that
low frequency breaks at hot-spots are more likely related to the ratio
between  the  outflow  distance   and  the  outflow  velocity  of  the
post-shocked material  after the Mach disk rather  than to synchrotron
aging.  Thus, we  will not use the information  of the break frequency
at the source extremes to derive synchrotron ages.

Middle panels in Fig.~\ref{spectra} correspond to the spectra taken at
711 kpc  (N-Lobe) and  903 kpc (S-Lobe)  from the  core, respectively.
Our data do not allow  us to distinguish between the Jaffe-Perola (JP;
Jaffe  \&  Perola   \cite{jaffe})  or  the  Kardashev-Pacholczyk  (KP;
Kardashev \cite{kardashev},  Pacholczyk \cite{pacholczyk}) synchrotron
loss models, both producing  equivalent results. An injection spectral
index  $\alpha_{inj} = 0.7$  has been  found and  kept fixed  in these
fits.   The discrepancy between  the injection  spectral index  in the
lobes  and in  the  hot-spots is  similar  to that  found in  Cygnus~A
(Carilli et al.~\cite{carilli}), although a physical interpretation of
this fact remains  unclear.  The derived break frequencies  are 15 GHz
at 711 kpc in the N-lobe and 18 GHz at 903 kpc in the S-lobe.

Bottom panels  in Fig.~\ref{spectra} refer to the  two regions nearest
to the  core, at 363 kpc  (N-Lobe) and 553 kpc  (S-Lobe).  Again, fits
using KP and  JP models are indistinguishable. The  break frequency is
1.6 GHz  in the N-lobe and  2.2 GHz in  the S-lobe.  Due to  the sharp
cut-off, the  flux densities  at 4.8  and 10 GHz  are below  the noise
level. In the fit we kept $\alpha_{inj} = 0.7$ as a fixed parameter.

\begin{figure}
\vspace{7cm}
\includegraphics{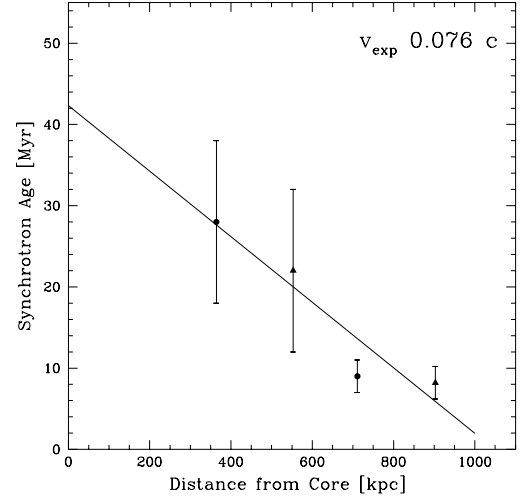}
%\rule{0.4pt}{4cm}
\caption{Spectral ages as a function of core distance in 8C\,0821+695. 
Filled dots refer to  the N-lobe; triangles to the  S-lobe. A mean expansion
velocity of 0.08 c is derived from a linear fit to the data. A source
spectral age of 42 Myrs is derived from the extrapolation at zero distance.} 
\label{age}
\end{figure}

Synchrotron ages were derived from the break frequencies in the lobes 
using the equation (Carilli et al.~\cite{carilli}):
\begin{equation}
t_{syn}=1.61\times 10^3 \frac{\sqrt{{\rm B}_{\rm eq}}}{{\rm B}_{\rm eq}^2 +
{\rm B}_{\rm IC}^2} \frac{1}{\sqrt{\nu_{\rm B}(1+{\rm z})}}.
\end{equation}
The synchrotron age $t_{syn}$ is  given in Myr, the break frequency
$\nu_{\rm B}$ in  GHz and magnetic fields in $\mu$G. For the strength of  
the equipartition magnetic
field   B$_{\rm   eq}$  we   took   the   corresponding  values   from
Tab.~\ref{param},  and  a magnetic  field  equivalent  to the  Inverse
Compton microwave  background of ${\rm  B}_{\rm IC} =  7.7\mu$G (${\rm
B}_{\rm IC} = 3.25 (1+{\rm z})^2 $). 
The spectral ages derived from the data are plotted in Fig.~\ref{age}.
The errors in the spectral ages 
can be derived from the uncertainties of ${\rm B}_{\rm eq}$ and 
$\nu_{\rm B}$; however, when 
${\rm B}_{\rm eq} \sim \frac{{\rm B}_{\rm  IC}}{\sqrt{3}}$ (as  it is
approximately given in our case), the total  error  is dominated by  the
uncertainties of  the break frequencies.
Therefore errors in Fig.~\ref{age} depend directly on the errors in the break 
frequencies which we have estimated considering the 1-$\sigma$ region
of allowance in the space of free-parameters in the fits. 
A weighted least-square fit yields a mean expansion velocity
of 0.08  c for both lobes, which  represents a measure of  the rate of
separation of the jet-head from the lobe material. The source spectral
age, obtained by extrapolation of the  age profiles up to the core, is
42 Myr. We note that the derived expansion velocity is consistent with that 
obtained assuming a zero age at the hot-spots, supporting the reliability of 
our spectral fit argument.

\section{Discussion}

\subsection{The ``conical'' lobe}

\begin{figure}
\vspace{9cm}
\includegraphics{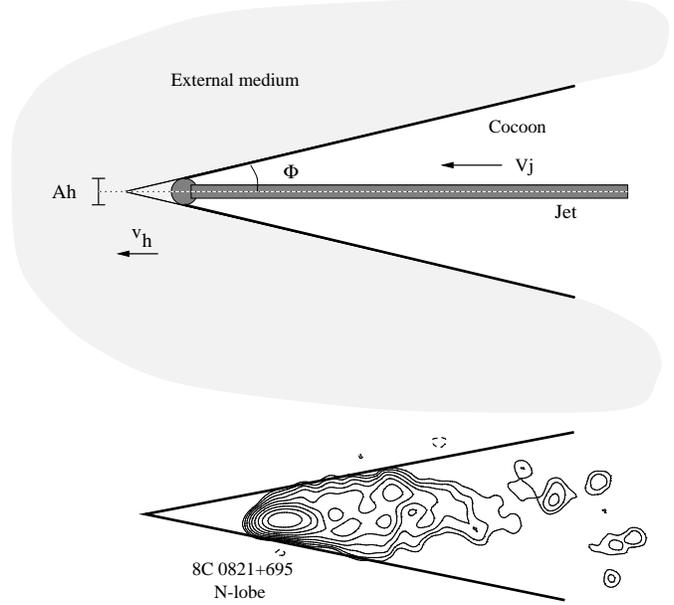}
\caption{Top: schematic representation of the jet and the lobe of a radio 
galaxy expanding in an uniform and homogeneous environment. 
Bottom: translation of this scheme to the northern lobe of 8C\,0821+695. 
\label{dibu}}
\end{figure}

In this section we propose a simple scenario of a jet nourishing a radio 
lobe with the  aim of 
constraining the  physical parameters of  the
jet itself and of the  ambient medium surrounding the radio emission.  We
consider the geometry outlined in Fig.~\ref{dibu}  for the jet and the
lobe of  a radio galaxy, and assume that
the external medium is  uniform and steady.  Relativistic  corrections
are not taken into account here, since lobe propagation velocities are
small compared with the speed of light (eg. Begelman et al. \cite{begelman2}).

In general, the ratio of the advance speed of the
emitting body (the jet-head velocity $v_{h}$) to the 
sound speed of the unperturbed external region ($v_{s}$), is 
related to the angle between the shock front and the velocity direction 
(the Mach angle $\Phi$), through the equation:
\begin{eqnarray}
\sin\Phi & = & \frac{v_{s}}{v_{h}} \nonumber \\ 
         & = & \frac{1}{v_{h}}\sqrt{\frac{\gamma P_{a}}{\rho_{a}}},
\label{1}
\end{eqnarray}
where $\gamma$, $P_{a}$ and $\rho_{a}$ are the adiabatic index, the pressure 
and the mass density of the ambient medium, respectively.
The mass density may also be written $\rho_{a}=\sigma n_{a}$, 
where $\sigma$ is
the mean atomic weight and $n_{a}$ is the number of particles 
per unit volume. Assuming an ideal gas, the pressure, density and temperature
are related through the equation of state $P_a = n_a k T_a$, where $k$ is the 
Boltzmann's constant and $T_a$ is the temperature of the ambient medium. 
Equation~\ref{1} provides a relationship between
$v_h$, the lobe geometry and the properties of the external medium. 

We note that $v_h$ as  given in Eq.~\ref{1} corresponds to the advance
velocity of  the jet-head with respect  to the external  medium at the
time of  the observations  since it depends  only on  local conditions
measured  close to the  hot-spot.  On  the other  hand, determinations
based  on  aging  arguments  (Sect.~3.4) correspond  to  the  jet-head
velocity  measured with  respect  to the  lobe  emitting material  and
averaged over the whole life of the radio source ($\langle v'_{h}
\rangle$).   Even if all our assumptions are correct, these two  
estimations   of   the    advance   velocity,   $v_h$   and   $\langle
v'_{h}\rangle$, may be different {\em  i)} if the backflow $v_{bf}$ of
the lobe  material is not negligible  or {\em ii)} if  the jet-head is
accelerated, so  that comparing  mean and instantaneous  velocities is
meaningless.  Besides,  incorrect  assumptions (e.g.  deviations  from
minimum energy conditions, a break frequency distribution not related to
the separation  velocity, wrong  estimations of the  external physical
parameters,  etc.)  may lead  to  differing  results  (see Carilli  et
al.  \cite{carilli} for a  detailed discussion).  In the  following we
will consider  that all our assumptions  are reasonably good
approximations to the real physical situation.

The  absence  or  presence  of  diffuse  extended  tails  of  emission
perpendicular to  the source axis provide information  of how important
the backflow of lobe material is.  In general,  we will consider that the 
mean backflow
velocity  is a  factor  $\epsilon$  of the  mean  advance velocity  of
the jet-head with respect to  the external medium, i.e. $\langle v_{bf} \rangle
= \epsilon \langle v_h \rangle$, so that the velocity of the head with
respect to the lobe $\langle v'_{h} \rangle = (1+\epsilon) \langle v_h
\rangle$. Similarly, the spectral age is related to the dynamical age of
the  source as $t=(1+\epsilon)  t_{syn}$.  Moreover,  we can  assume a
constant jet-head deceleration and determine the initial head velocity
$v_{hi}$ as
\begin{equation}
v_{hi}=2(1+\epsilon)^{-1} \langle v'_{h} \rangle -  v_{h}. 
\label{2}
\end{equation}

On the other hand, the 
balance of the ram pressure of the ambient medium and the thrust of the jet, 
spread over the cross-sectional area $A_{h}$ of the bow
shock at the end of the jet is (Begelman \& Cioffi 1989):
\begin{equation}
v_{h}\sim\sqrt{\frac{L_{j}}{\sigma~n_{a}v_{j}A_{h}}},
\label{3}
\end{equation}
where  $L_j$  is  the   total jet power  and  $v_j$  is the  jet  bulk
velocity. In  an uniform medium and  if these two  jet parameters were
constant  with time,  a  varying  $v_h$ could  be  obtained through  a
variation  in the  contact surface  $A_h$.   
We note that a constant $A_{h}$ and $n_{a}$ decreasing  
with  core distance would  lead to $v_{h}$ increasing with time, 
a situation highly  unplausible (Loken et al. \cite{loken}).  Although  a  
combination  of  both
situations might occur  leading  to a given $v_{h}$ evolution,
we  will neglect density variations for simplicity.
Therefore, from 
Eq.~\ref{3}, we find a relation between $v_h$ and $A_h$ at present
and initial conditions:
\begin{equation}
A_h = A_{hi} \left( \frac{v_{hi}}{v_h} \right )^2 .
\label{4}
\end{equation}

This relation translates to an opening angle ($\theta$),
defined by the time evolution of the contact surface, given by
\begin{equation}
\tan \theta = \frac{\sqrt{\frac{A_h}{\pi}} 
\left(1- \frac{v_{h}}{v_{hi}} \right )}{l},
\label{5}
\end{equation}
where $l$ is the total jet length.

\subsection{Application to 8C\,0821+695}

The  N-lobe  of 8C\,0821+695  clearly  resembles  the simple  geometry
outlined  in Fig.~\ref{dibu},  so  it seems  reasonable  to apply  the
previous  analysis to  this lobe. Even if we need to make several 
assumptions and the uncertainties of our results are high, we obtain at least 
an idea of the order of magnitude of the different parameters, which is the
aim of these calculations.

From the VLA maps at  4860 and 1425 MHz (Fig.~\ref{vla}), we measure
an  angle  $\Phi=12^{\circ}$  at  the  N-lobe  (see  Fig.~\ref{dibu}).
However, we  note that this  value is a  lower limit to the  true Mach
angle  since we  implicitly assume  that the  edge of  the synchrotron
emitting region defines the lobe contact discontinuity, and that this
discontinuity is coincident with the
bow  shock   produced  by  the  jet-head  advance   in  the  external
medium.  Such  assumption  underestimates  the true  Mach angle,  since  it
implies that the region of shocked external medium between  the contact 
discontinuity and the bow shock is negligible. Moreover,   we  assume   
$\sigma=1.4$~amu,
$\gamma=\frac{5}{3}$  and  an  upper  limit of  $T_a=10^7$~K  for  the
external medium temperature (Barcons et al. \cite{barcons}).

We have accepted the minimum energy conditions and used the NVSS and  WENSS 
maps to derive the  cocoon pressure at low brightness  regions in the
lobes of 8C\,0821+695, far away from the hot-spot; we obtain $P_c
\sim 1.5\times 10^{-14}$ Nm$^{-2}$.  Strictly speaking, this value
constitutes an  upper limit to  the external pressure $P_a$  since the
bridges in  FR II radio  sources are most probably  overpressured with
respect to the surrounding medium (e.g.  Subrahmanyan \& Saripalli
\cite{subra1}; Nath \cite{nath}). However, observations (Subrahmanyan
et al.  \cite{subra2}) and  models (Kaiser \& Alexander \cite{kaiser})
indicate that the lobes of FR II radio galaxies grow in a self-similar
way, so  that GRGs  are expected to  have the lowest  pressures, being
closer to  a situation of  pressure equilibrium with the  outer medium
than other FR  II radio galaxies.  Thus, we  can reasonably assume for
our calculations  that the low-brightness  regions in
the lobes of 8C\,0821+695 provide a good approximation of the ambient
gas pressure  $P_{a}$ (see also Schoenmakers et al. \cite{arno2}). In fact, 
the  pressure we obtain is  as low as
that found at  the very periphery of galaxy  clusters, confirming that
the environment of this giant source is tenuous. 

Considering the  upper limit temperature  of $10^7$~K for  the ambient
gas,  the particle  density resulting  from the  equation of  state is
about $10^2$~m$^{-3}$.   This density is consistent with  the limit on
$RM$  ($|RM| \le  20$ rad  m$^{-2}$) if  the ambient  uniform magnetic
field is about 0.5 $\mu$G and the Faraday depth less than 1 Mpc, which
are reasonable  values since  we do  not expect a  strong field  to be
distributed over large regions.

From Eq.~\ref{1},  we obtain  $v_h \sim 5\times  10^{-3}$~c which, 
according to our assumptions is an
upper limit to the true jet-head velocity.  Besides,
from spectral aging arguments (Sect.~3.4) we obtained a mean expansion
velocity of the head with respect to the lobe material $\langle v'_{h}
\rangle = 0.08$~c.  On the other hand, Lacy et al. (\cite{lacy}) find  
evidence of a tail at the  base of the N-lobe and suggest a backflow speed  
close to the advance speed of the head. According to that, we will  
assume $\epsilon \simeq 1$, so
that  $\langle  v_{h} \rangle  \simeq  \langle  v_{bf} \rangle  \simeq
0.04$~c.  Using Eq.~\ref{2}, we  estimate an initial jet-head velocity
$v_{hi} \sim 0.075$~c.

The area of the head contact surface $A_{h}$ can be derived by fitting
an elliptical Gaussian  to the northern hot-spot at 5  GHz. We obtain a
deconvolved angular  diameter of  $\sim 5\arcsec$  ($\sim 25$  kpc).  
We  can then
calculate the total  power of the jet from  Eq.~\ref{3}, assuming a
jet bulk velocity $v_j \sim$c (Fernini et al. \cite{fernini}).  
We obtain $L_j = 7.1\times 10^{37}$
W.  Alternatively,  we can estimate  the total power from  the average
minimum energy  density ($\bar{u}_{me}=2.7\times 10^{-13}$ J m$^{-3}$; 
Tab.~\ref{param}),  the volume of  the source
(simplified to a cylinder of 2~Mpc$\times$200~kpc)  and its dynamical age 
$t=2 t_{syn} = 84$ Myr, giving
$L = 9\times 10^{37}$W, fully consistent with the previous result. 

From Eq.~\ref{4}, the  angular diameter of the jet contact surface at the
initial  stages  of  the  source  is  about  0\farcs33.  The
increase  in the contact  surface  from this value to the measured one 
($5\arcsec$) requires an  opening  angle $\theta$, defined by the hot-spot 
size evolution, of  0\fdg7 (Eq.~\ref{5}).
Thus, a small increase of the hot-spot size with time can naturally 
explain the  deceleration of the
jet-head and the difference  in the velocity determinations from local
and global conditions.

Finally, the radio power of the northern lobe $L_{r}$ can be derived from the 
observed flux density and spectral index:
\begin{equation}
L_{r}=4\pi D_{L}^{2}\int_{\nu_{1}}^{\nu_{2}}S(\nu)d\nu ,
\end{equation}
where $D_{L}$ is the luminosity distance, and $\nu_{1}=10$ MHz and 
$\nu_{2}=100$ GHz are
the assumed lower and upper frequency cutoffs. Since $S(\nu)\propto 
\nu^{-\alpha}$ and $\alpha=1.1$, we obtain
$ L_{r}=6.9\times 10^{35}$ W. $L_{r}$ being about two orders of magnitude 
lower than $L_{j}$ would indicate that most of the jet power is devoted 
to the expansion of the lobe against the external medium. Results are 
summarized in Table~\ref{param2}.

\noindent

\begin{table}[h]
\caption[]{8C\,0821+695 and its environment}
\label{param2}
\begin{tabular}{rrll} 
\hline 
Ambient density         & $n_{a} $ & $\sim 100$              & m$^{-3}$  \\
Ambient pressure        & $P_{a} $ & $\sim 1.5\times 10^{-14}$ & N m$^{-2}$\\ 
Present head velocity   & $v_{h} $ & $\sim 0.005$            & c         \\
Mean head velocity      &$\langle v_h \rangle$ & $\sim 0.04$ & c         \\ 
Initial head velocity   & $v_{hi}$ & $\sim 0.075$            & c         \\
Spectral Age            & $t_{syn}$& $\sim 4.2\times 10^{7}$ & yr        \\
Dynamical Age           & $t $     & $\sim 8.4\times 10^{7}$ & yr        \\ 
Northern jet Power      & $L_{j}$  & $\sim 7\times 10^{37}$  & W         \\ 
N-lobe radio power      & $L_{r}$  & $\sim 7\times 10^{35}$  & W         \\ 
Opening angle           & $\theta$ & $\sim 0\fdg7$         &    \\ 
\hline 
\end{tabular}
\end{table}

The  S-lobe  does not  present  such  a  suitable morphology  for  the
application of  the previous simple model  since the head  of the lobe
does  not have  a clear  cone-like appearance (see Sect.~2.3).  
From the observed arm-ratio, we might deduce that the external medium here
could be more tenuous than in the northern lobe region.

\section{Conclusions}

We  present new radio  observations made  with the  VLA and  the 100-m
Effelsberg  radio  telescope, of  the  GRG  8C\,0821+695 at  different
frequencies  and  angular resolutions.  Our  data  have been  analyzed
together with survey and literature data in order to study the details
of  a high-redshift  GRG, and  obtain information  about  the external
medium surrounding the radio source.

8C\,0821+695  is a  straight 2  Mpc long  FR II  radio source,  with a
north-to-south  arm-ratio of  0.89.  The  N-lobe contains  a hot-spot,
responsible  for most  of the  source polarized  emission.  The S-lobe
presents a  more relaxed structure,  without a well  defined hot-spot.
At  high   frequencies  ($\nu\geq  1400$~MHz)   8C\,0821+695  shows  a
prominent  compact core.  We  have not  found  any trace  of the  jets
nourishing the lobes.

The  spectral index  distribution over  the lobes  of  8C\,0821+685 is
typical  of FR II-type  radio galaxies,  showing a  gradual steepening
from the outer ends towards the core.  The mean lobe spectral index is
$\alpha=1.1$.  The core has a flat spectrum with $\alpha= 0.3$.  Using
the  available data,  we have  made a  spectral-aging analysis  of the
source lobes, providing the dependence of the spectral break frequency
and the synchrotron  age with the distance from the  core. We obtain a
mean  expansion velocity  of the  jet head  with respect  to  the lobe
material of $0.08$~c, and a spectral  age of 42 Myr. This age determination
might be affected by the possible existence of backflow of material
in the lobe, being a lower limit to the true age. The age we derive for 
8C\,0821+695 is  of the order  of ages  estimated for  other GRGs  
(Schoenmakers et al. \cite{arno2}). 

We have studied  the $RM$ over the lobes  of 8C\,0821+695, obtaining a
smooth and uniform distribution,  which we ascribe to Faraday rotation
mostly produced by the Galactic medium.  Equipartition conditions have
been assumed  in order to derive  physical parameters of  the lobes at
different  positions, yielding magnetic  fields, pressures  and energy
densities that are consistent with estimated values in other GRGs.

Under very simple assumptions we have estimated physical parameters of
the jet  and the  external medium of  8C\,0821+695.  We find  that the
present  expansion  velocity  is  significantly lower  than  the  mean
expansion velocity even if backflow is allowed, implying the existence
of deceleration.  We  explain this deceleration by an  increase of the
cross-sectional  area of  the bow  shock at  the end  of the  jet with
time. We  note that external  density estimates in the  literature for
other GRGs usually  consider the contact surface measured  at the time
of the observations together  with mean quantities (like the expansion
velocity  derived   from  aging  arguments),   resulting  in  external
densities lower than the density  we obtain.  However, our results are
still consistent with GRGs evolving in poor density regions.

\begin{acknowledgements}

We  thank the  referee  Dr.  R. Perley  for  helpful and  constructive
comments to the paper.  The  National Radio Astronomy Observatory is a
facility of the National Science Foundation operated under cooperative
agreement by Associated Universities, Inc.  This research has made use
of the NASA/IPAC Extragalactic Database (NED) which is operated by the
Jet Propulsion  Laboratory, California Institute  of Technology, under
contract with the National  Aeronautics and Space Administration. This
research is supported  in part by the Spanish  DGICYT (PB97-1164). KHM
was  supported by  the  European Commission,  TMR Programme,  Research
Network Contract ERBFMRXCT96-0034 ``CERES''. LF, GG and MM acknowledge
a partial support by the  Italian Ministry for University and Research
(MURST) under grant Cofin98-02-32.

\end{acknowledgements}


\begin{thebibliography}{}

\bibitem[1977]{baars} Baars J.M.W., Genzel R., Pauliny-Toth I.I.K., Witzel A.,
1977, \aa, 61, 99

\bibitem[1991]{barcons} Barcons X., Fabian A.C. \& Rees M.J., 1991, Nature, 
350, 685

\bibitem[1984]{begelman2} Begelman M.C., Blandford R.D., Rees M.J., 1984, Rev. Mod. Phys., 56, 255

\bibitem[1989]{begelman} Begelman M.C., Cioffi D.F., 1989, \apj, 345, L21

\bibitem[1991]{carilli} Carilli C.,  Perley R.A., Dreher J.H., Leahy J.P., 
1991, \apj, 383, 554

\bibitem[1998]{nvss} Condon J.J., Cotton W.D., Greisen E.W., Yin Q.F.,
Perley R.A., Taylor G.B., Broderick J.J., 1998, \anj, 115, 1693

\bibitem[1974]{fanaroff} Fanaroff B.L., Riley J.M., 1974, \mn, 167, 31

\bibitem[1997]{fernini} Fernini I., Burns J.O., Perley R.A., 1997, \anj, 
114, 2292 

\bibitem[1992]{gregorini} Gregorini L., Klein U., Parma P., Wielebinski R., 
Schlickeiser R., 1992, \aasup, 94, 13 

\bibitem[1999]{ishwara} Ishwara-Chandra C.H., Saikia D.J., 1999, \mn, 309, 100 

\bibitem[1973]{jaffe} Jaffe W.J., Perola G.C., 1973, \aa, 26, 423

\bibitem[1997]{kaiser} Kaiser C.R., Alexander P., 1997, \mn, 286, 215

\bibitem[1962]{kardashev} Kardashev, N.S., 1962, \anj, 6, 317

\bibitem[1986]{killeen} Killeen N.E.B., Bicknell G.V., Ekers R.D., 1986,
\apj, 302, 306

\bibitem[1995]{kleinmack} Klein U., Mack K.-H., 1995, Proceedings Workshop
on  ``Multi-Feed    Systems    for   Radio    Telescopes'',    Tucson,
Ed. D.T. Emerson, ASP Conference Series 

\bibitem[1993]{lacy} Lacy M., Rawlings S., Saunders R., Warner P.J., 1993, 
\mn, 264, 721

\bibitem[1999]{lara} Lara L., M\'arquez I., Cotton W.D., Feretti L., Giovannini G., Marcaide J.M., Venturi T., 1999, \nar, 43, 643

\bibitem[1992]{loken} Loken C., Burns D.O., Clarke D.A., Norman M.L., 1992, \apj, 392, 54

\bibitem[1997a]{mack97a} Mack K.-H., Kerp J., Klein U., 1997a, \aa, 324, 870

\bibitem[1997b]{mack97b} Mack K.-H., Klein U., O'Dea C.P., Willis A.G., 
1997b, \aa, 123, 423

\bibitem[1989]{meisenheimer} Meisenheimer K., R\"oser H.-J., Hiltner P.R., 
Yates M.G., Longair M.S., Chini R., Perley R.A., 1989, \aa, 219, 63 

\bibitem[1980]{miley} Miley G., 1980, \rev, 18, 165

\bibitem[1995]{nath} Nath B.B., 1995, \mn, 274, 208

\bibitem[1970]{pacholczyk} Pacholczyk, A.G., 1970, Radio Astrophysics (San Francisco: Freeman)

\bibitem[1996]{parma} Parma P., de Ruiter H.R., Mack K.-H., van Breugel W., 
Dey A., Fanti R., Klein U., 1996, \aa, 311, 49

\bibitem[1990]{rees} Rees N., 1990, \mn, 244, 233

\bibitem[1980]{simard} Simard-Normandin M., Kronberg P.P., 1980, \apj, 242, 74

\bibitem[1997]{wenss} Rengelink R., Tang Y., de Bruyn A.G., Miley G.K., 
Bremer M.N., R\"ottgering H.J.A., Bremer M.A.R., 1997, \aasup, 124, 259 

\bibitem[1982]{scheuer} Scheuer P.A.G., 1982, in Heeschen D.S., Wade C.M. eds,
Proc.    IAU Symp.   97,  ``Extragalactic Radio   Sources''.   Reidel,
Dordrecht, p.163 

\bibitem[2000a]{arno1} Schoenmakers A.P., de Bruyn A.G., R\"ottgering H.J.A.,
van der Laan H., 2000a, \aa, in preparation

\bibitem[2000b]{arno2} Schoenmakers A.P., Mack K.-H., de Bruyn A.G.,
R\"ottgering H.J.A., Klein U., van der Laan H., 2000b, \aa, in press

\bibitem[1993]{subra1} Subrahmanyan R., Saripalli L., 1993, \mn, 260, 908

\bibitem[1996]{subra2} Subrahmanyan R., Saripalli L., Hunstead R.W., 1996, 
\mn, 279, 257

\bibitem[1974]{wardle} Wardle J.F.C., Kronberg P.P., 1974,
\apj, 194, 249

\bibitem[1991]{wieringa} Wieringa M., 1991, PhD. Thesis, Univ. Leiden

\bibitem[1990]{willis} Willis A.G., O'Dea C.P., 1990, in ``Galactic and 
Intergalactic Magnetic Fields'' IAU Symp. 140, Beck R., Kronberg P.P. \& 
Wielebinski R. (eds.), Reidel, Dordrecht, p.455


\end{thebibliography}
\end{document}